\documentclass[journal]{IEEEtran}
\usepackage{nccmath}
\usepackage{amssymb}
\usepackage{cite}
\usepackage{graphicx}

\ifCLASSINFOpdf
\else
\fi

\newcommand{\argmax}{\mathop{\rm arg\ max}\limits}

\begin{document}
%
\title{A Scale Mixture-based Stochastic Model of\\ Surface EMG Signals with Variable Variances}
%
%
%

\author{Akira~Furui,~\IEEEmembership{Student Member,~IEEE,}
        Hideaki~Hayashi,~\IEEEmembership{Member,~IEEE,}
        and~Toshio~Tsuji,~\IEEEmembership{Member,~IEEE}
\thanks{A.~Furui and T.~Tsuji are with the Graduate School of Engineering,
Hiroshima University, Higashi-hiroshima, 739-8527 Japan (e-mail: akirafurui@hiroshima-u.ac.jp).}
\thanks{H.~Hayashi is with the Department of Advanced Information Technology,
Kyushu University.}
\thanks{Copyright (c) 2019 IEEE. Personal use of this material is permitted. Permission from IEEE must be obtained for all other uses, in any current or future media, including reprinting/republishing this material for advertising or promotional purposes, creating new collective works, for resale or redistribution to servers or lists, or reuse of any copyrighted component of this work in other works.}
\thanks{This paper is an accepted version for publication in IEEE Transactions on Biomedical Engineering. The final version is available at {https://ieeexplore.ieee.org/document/8627996}.}}

\maketitle

\begin{abstract}
\textit{Objective:}
Surface electromyogram (EMG) signals have typically been assumed to follow a Gaussian distribution.
However, the presence of non-Gaussian signals associated with muscle activity has been reported in recent studies, and there is no general model of the distribution of EMG signals that can explain both non-Gaussian and Gaussian distributions within a unified scheme.
\textit{Methods:}
In this paper, we describe the formulation of a non-Gaussian EMG model based on a scale mixture distribution.
In the model, an EMG signal at a certain time follows a Gaussian distribution, and its variance is handled as a random variable that follows an inverse gamma distribution.
Accordingly, the probability distribution of EMG signals is assumed to be a mixture of Gaussians with the same mean but different variances.
The EMG variance distribution is estimated via marginal likelihood maximization.
\textit{Results:}
Experiments involving nine participants revealed that the proposed model provides a better fit to recorded EMG signals than conventional EMG models.
It was also shown that variance distribution parameters may reflect underlying motor unit activity.
\textit{Conclusion:}
This study proposed a scale mixture distribution-based stochastic EMG model capable of representing changes in non-Gaussianity associated with muscle activity.
A series of experiments demonstrated the validity of the model and highlighted the relationship between the variance distribution and muscle force.
\textit{Significance:}
The proposed model helps to clarify conventional wisdom regarding the probability distribution of surface EMG signals within a unified scheme.
\end{abstract}


%
\IEEEpeerreviewmaketitle

\section{Introduction}
%
%
%
%
\IEEEPARstart{S}{urface} electromyogram (EMG) signals are the summation of individual action potentials generated from motor units.
EMG signals are recorded from the skin surface and reflect the internal state of muscle activity.
As they can be recorded noninvasively, these signals have been applied in a wide range of fields, including motion analysis~\cite{Moynes1986,Millington1992,tao2012}, neuromuscular system analysis~\cite{Baweja2009,Ganguly2011,Busche2015}, neuromuscular disease diagnosis~\cite{Sturman2005,Cifrek2009,Dideriksen2011}, and prosthesis control~\cite{Fukuda2003,Scheme2011,Farina2017}.

In general, EMG signals are treated as a stochastic process because individual motor unit firing activities can be considered probabilistic events.
There have been many attempts to extract their features by modeling the stochastic properties of EMG signals, which are typically assumed to follow a Gaussian distribution~\cite{Parker1977,Hogan1980c,Abbink1998}.
Parker \textit{et al.} experimentally showed that measured EMG signals can be fitted using a Gaussian distribution~\cite{Parker1977}.
Hogan and Mann modeled the relationship between EMG signals and muscle force based on a Gaussian distribution~\cite{Hogan1980c}.
They assumed that the EMG variance was constant under a constant-force condition and estimated the variance using the maximum-likelihood method.

EMG signals, however, do not always follow a steady Gaussian distribution, even under constant-force conditions~\cite{Milner-Brown1975,Hunter1987,Clancy1999,Bilodeau1997,Naik2011,Nazarpour2013}.
Milner-Brown and Stein showed that the distributions of EMG signals recorded from the first dorsal interosseous (FDI) muscle under a constant-force condition are sharper than a Gaussian distribution and peak near zero~\cite{Milner-Brown1975}.
The same tendency was observed in EMG signals recorded from the biceps brachii (BB) muscle~\cite{Hunter1987}.
Clancy and Hogan experimentally observed that the probability distribution of EMG signals falls between a Gaussian and a Laplacian distribution~\cite{Clancy1999}.
To analyze the non-Gaussianity of EMG signals, high-order statistics, such as the kurtosis and bispectrum approaches, have recently been used~\cite{Nazarpour2013, Naik2011, Zhao2012,Messaoudi2017}.
This approach revealed that EMG signals have a heavier-tailed distribution than in the Gaussian case at low contraction levels~\cite{Nazarpour2013, Naik2011}.
Some simulation studies have also reported that the non-Gaussianity of EMG signals varies depending on the muscle contraction level such that the increase in the contraction level shifts the probability distribution of EMG signals towards the Gaussian distribution~\cite{Zhao2012,Messaoudi2017}.
Despite these experimental reports, there is no general consensus on the distribution of EMG signals, and a stochastic model that can represent the non-Gaussianity of EMG signals depending on muscle activity has not been developed.

Hayashi \textit{et al.} focused on the uncertainty of EMG variance (i.e., EMG amplitude) and proposed an EMG signal model under the assumption that the variance is a random variable~\cite{Hayashi2017}.
In this variance distribution model, the EMG signals follow a non-Gaussian distribution with a heavy tail because the variance of EMG signals at a certain time is randomly determined.
However, this model does not allow a wide-ranging variance distribution because one variance distribution parameter is fixed to enable real-time estimation.
In addition, the relationship between the muscle contraction level and variance distribution as well as the goodness-of-fit of the model to real EMG signals have not been revealed.
If we could identify these, a novel stochastic modeling framework for EMG signals would be established.

Here, we interpret the variance distribution model as a scale mixture model for the variance and outline the estimation and analysis of the EMG variance distribution.
In the proposed scale mixture model, an EMG signal at a certain time follows a Gaussian distribution, and its variance is handled as a random variable that follows an inverse gamma distribution.
The EMG variance distribution based on such a relationship is estimated via marginal likelihood maximization without fixed distribution parameters.
This allows an estimation to determine the variance distribution parameters where the observed EMG signals are most likely to occur.
In the experiments, the goodness-of-fit of the model is evaluated using EMG signals measured from FDI and BB muscle.
The relationship between the variance distribution and muscle activity is also examined through the change in variance distribution parameters according to muscle force.

The reminder of this paper is organized as follows.
Section~II outlines the structure of the scale mixture model and the parameter estimation method, before Section~III details the experimental setup for the model verification and EMG analysis experiments.
Section~IV presents the results of these experiments, and Section~V provides related discussion.
Finally, Section~VI states the conclusions to this study.

\section{Scale Mixture Model of Surface EMG Signals}

\subsection{Model Structure}
\begin{figure}[!t]
\centering
\includegraphics[width=1.0\hsize]{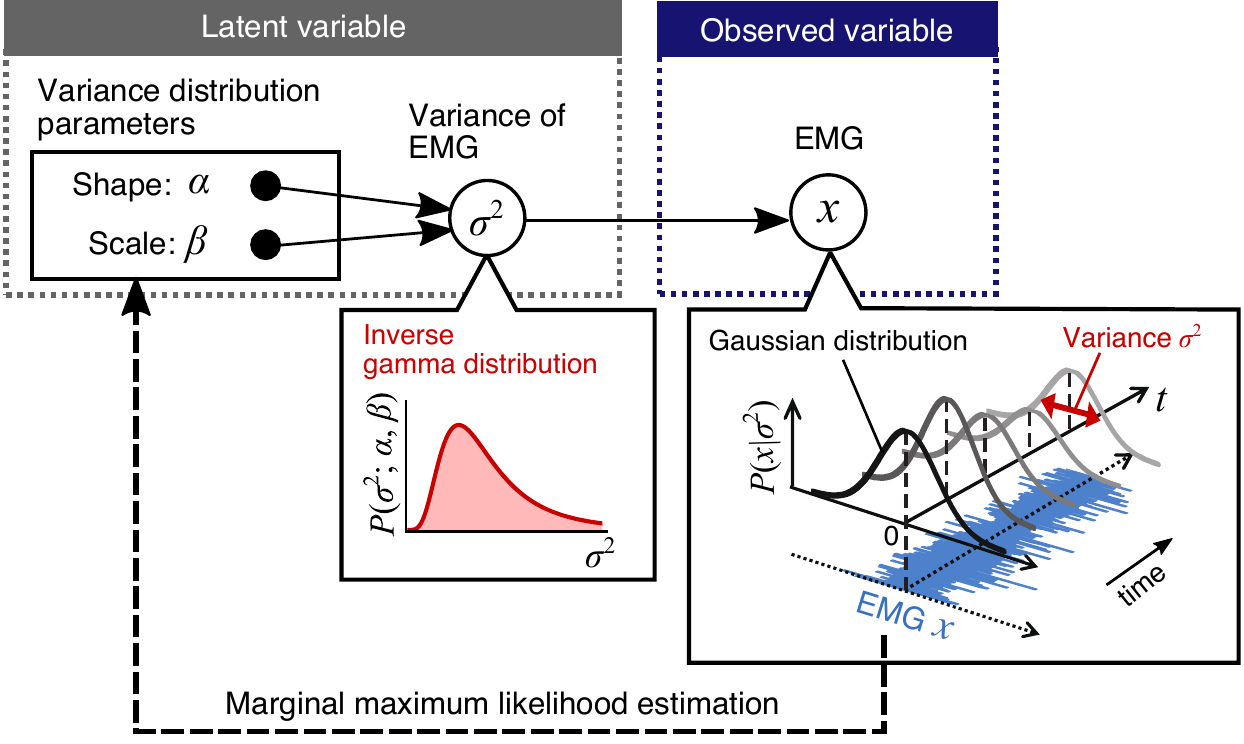}
\caption{Graphical representation of the stochastic relationship between the EMG signal and its variance.
The white nodes are random variables, and the black nodes are parameters to be estimated.
In the model, the EMG signal $x$ is handled as a random variable that follows a Gaussian distribution with a mean of zero.
The EMG variance $\sigma^2$ is also a random variable that follows the inverse gamma distribution determined by the shape parameter $\alpha$ and the scale parameter $\beta$.
The variance distribution parameters are estimated via marginal likelihood maximization from measured EMG signals.}
\label{fig:model}
\end{figure}

Fig.~\ref{fig:model} shows the stochastic relationship between the EMG signal $x$ and its variance $\sigma^2$ as a graphical model.
$x$ is handled as a random variable that follows a Gaussian distribution with a mean of zero and a variance of $\sigma^2$.
The variance is also a random variable for which the distribution is determined by the shape parameter $\alpha$ and the scale parameter $\beta$.
In the model, the variance $\sigma^2$ is interpreted as a latent variable because it is not directly observed.
Note that the frequency components of EMG signals are ignored in the model; only their variance is considered.

First, the conditional distribution of an EMG signal $x$ given $\sigma^2$ is expressed via the following Gaussian distribution:
\begin{equation}
	P(x|\sigma^2) = {\mathcal N}(x|\sigma^2) = \frac{1}{\sqrt{2\pi\sigma^2}} {\rm exp} \left[-\frac{x^2}{2\sigma^2}\right].
\label{eq:p_x_sigma}
\end{equation}
Considering that $\sigma^2 > 0$, the variance is assumed to obey an inverse gamma distribution ${\rm IG}(\sigma^2; \alpha, \beta)$, which is known as a conjugate distribution for the variance of a Gaussian distribution~\cite{Hayashi2017,Furui2017a}:
\begin{equation}
	P(\sigma^2) = {\rm IG}(\sigma^2;\alpha,\beta) = \frac{\beta^{\alpha}}{\Gamma (\alpha)}(\sigma^2)^{-\alpha-1} {\rm exp} \left[-\frac{\beta}{\sigma^2}\right]
\label{eq:p_sigma}
\end{equation}
where $\alpha$ and $\beta$ determine the inverse gamma distribution and are referred to as the shape parameter and the scale parameter, respectively.
Considering the marginal distribution of $x$, the variance $\sigma^2$ can be integrated out, as follows:
\begin{eqnarray}
	P(x) &=& \int P(x|\sigma^2)P(\sigma^2) {\rm d}\sigma^2\nonumber \\
		&=& \int {\mathcal N}(x|\sigma^2) {\rm IG}(\sigma^2;\alpha,\beta) {\rm d}\sigma^2 \label{eq:marginal_x} \\
		&=& \frac{\beta^{\alpha} \Gamma \left(\alpha + \frac{1}{2}\right) }
		{\sqrt{2\pi}\Gamma(\alpha) (\beta + \frac{x^2}{2} )^{\alpha+\frac{1}{2} } }.
\label{eq:p_x}
\end{eqnarray}
From equation (\ref{eq:marginal_x}), $P(x)$ is obtained by summing an infinite number of Gaussian distributions having the same mean but different variances.
This can be interpreted as an infinite mixture of Gaussian distributions for the scale parameter (i.e., variance).

Note that this model describes the macroscopic electrical activity of muscle.
Underlying structural processes, such as firing and the recruitment patterns of motor units, are not explicitly formulated, but the sum of such individual events is assumed to be included as part of the random processes.
Therefore, the EMG variance distribution and its parameters, which characterize the stochastic properties of EMG variance, may reflect the underlying characteristics of muscle activities.
Hayashi \textit{et al.} assumed that the shape parameter $\alpha$ remains constant regardless of muscle activities and introduced an estimation method for the variance distribution using rectified/smoothed EMG signals to enable real-time estimation~\cite{Hayashi2017}.
In this method, only the scale parameter $\beta$ is estimated.
In contrast, Furui \textit{et al.} examined the relationship between muscle force and the variance distribution parameters and indicated that both $\alpha$ and $\beta$ change according to muscle force.
Therefore, this paper estimates the variance distribution via the marginal likelihood maximization of EMG signals without fixed distribution parameters, thereby allowing a wide range of distribution shapes to be determined.
The next subsection outlines a method for the estimation of $\alpha$ and $\beta$ from EMG signals.

\subsection{Variance Distribution Estimation Based on Marginal Maximum Likelihood}

Let us consider the estimation of $\alpha$ and $\beta$, given $N$ samples of EMG signals $X = \{x_n\}^N_{n=1}$.
The parameters where observations are most likely to occur can be estimated by maximizing the marginal likelihood $P(X) = \prod^{N}_{n=1} P(x_n)$.
Hayashi \textit{et al.} maximized this using the steepest descent method~\cite{Hayashi2017}, but the convergence stability of this approach is sensitive to the initial values and step size.
This is because direct optimization of marginal likelihood is generally problematic~\cite{Bishop2006}.
Therefore, we conduct this optimization based on the expectation maximization (EM) algorithm~\cite{Furui2017a}, which is an effective approach for models having latent variables.
In this algorithm, marginal likelihood is maximized indirectly by maximizing the expectation of the complete-data log likelihood, as this is equivalent to maximizing the lower bound of the marginal likelihood~\cite{Bishop2006}.
The EM algorithm is iterated via application of an expectation step (E-step) and a maximization step (M-step).

Equation (\ref{eq:marginal_x}) is transformed via the introduction of the new parameterization $\nu = 2\alpha$ and $s = \beta/\alpha$, with the latent variable redefined as $\tau_n = \sigma^2_n \alpha/\beta$.
Marginal distribution can then be expressed as
\begin{eqnarray}
		P(x_n) &=& \int {\mathcal N}(x_n|s\tau_n) {\rm IG}(\tau_n;\nu/2,\nu/2) {\rm d}\tau_n \nonumber \\
		&=& \frac{\Gamma \left(\frac{\nu+1}{2}\right)s^{-\frac{1}{2}} }
		{\sqrt{\pi\nu}\Gamma(\frac{\nu}{2}) (1 + \frac{x_n^2}{\nu s} )^{\frac{\nu+1}{2} } }.
\label{eq:param_marginal_likelihood}
\end{eqnarray}
This expression is equivalent to Student's $t$-distribution with a mean of zero, in which $\nu$ is the degree of freedom and $s$ is the scale parameter of the $t$-distribution~\cite{Bishop2006}.
Here, the posterior distribution of the latent variable is given by
\begin{eqnarray}
	P(\tau_n|x_n) &\propto & {\mathcal N}(x_n|s\tau_n) {\rm IG}(\tau_n;\nu/2,\nu/2) \nonumber \\
	&\propto & {\rm IG}\left(\tau_n;\frac{\nu+1}{2},\frac{\nu+x_n^2/s}{2}\right).
	\label{eq:expectation_latent}
\end{eqnarray}
The parameters $[\nu, s]$ are estimated as outlined below based on the EM algorithm.
\begin{enumerate}
\renewcommand{\labelenumi}{(\roman{enumi})}
	\item Initialize $\nu$ and $s$ with the selection of arbitrary starting values.
	\item E-step.
	      Calculate the expectation of the complete-data log likelihood.
		  This expectation, denoted as $Q(\nu,s)$, is given by
	      \begin{fleqn}
	      \begin{eqnarray}
			Q(\nu,s) &=& \mathbb{E} \left[ \ln \prod_{n=1}^{N} {\mathcal N}(x_n|s\tau_n)
						 {\rm IG}(\tau_n;\nu/2,\nu/2)\right] \nonumber \\
					 &=& \sum_{n=1}^{N} \biggl\{ -\frac{1}{2} \ln (2\pi) -\frac{1}{2} 
						 \ln s - \frac{x_n^2}{2s} \omega_n + \frac{\nu}{2} \ln \frac{\nu}{2} \nonumber \\
					&\quad& - \ln \Gamma \left(\frac{\nu}{2} \right) - \frac{\nu}{2} \omega_n 
						 -\left(\frac{\nu}{2}+\frac{3}{2} \right) \lambda_n \biggr\} 
	      \end{eqnarray}
	      \end{fleqn}
		   where $\omega_n$ and $\lambda_n$ are derived as follows by calculating the posterior distribution of the latent variable from (\ref{eq:expectation_latent}).
		\begin{eqnarray}
			\omega_n &:=& \mathbb{E}\,[\tau^{-1}_n|x_n] = \frac{\nu+1}{\nu + x_n^2/s}, \\
			\lambda_n &:=& \mathbb{E}\,[\ln \tau_n|x_n] \nonumber \\
						&=& -\ln \omega_n + \ln \left( \frac{\nu + 1}{2} \right) - \psi \left( \frac{\nu + 1}{2} \right)
		\end{eqnarray}
		where $\psi(\cdot)$ is a digamma function.
	\item M-step. Update the parameters by maximizing $Q(\nu, s)$.
	By setting the derivative of $Q(\nu, s)$ with respect to $s$ equal to zero, the new scale parameter is obtained as
		\begin{equation}
			s^{\textrm{new}} = \frac{1}{N} \sum_{n=1}^{N} \omega_n x_n^2.
		\end{equation}
		There is no closed-form expression for the degree of freedom. Hence, it is updated by finding $\nu$, where $Q(\nu, s)$ is maximized numerically based on a line search.
		\begin{equation}
			\nu^{\textrm{new}} = \argmax_{\nu}\, Q(\nu,s^{\textrm{new}}).
		\end{equation}
	\item Evaluate the log-marginal likelihood
    \begin{equation}
      \ln P(X) = \ln \prod^{N}_{n=1} P(x_n)
    \label{eq:param_marginal_likelihood}
    \end{equation}
		and determine whether the convergence of the estimation is reached when the relative change in the log-marginal likelihood between iterations falls below the predetermined threshold $\varepsilon$.
		If the convergence criterion is not satisfied, return to step (ii).
\end{enumerate}

As each cycle of the EM algorithm will increase the log-marginal likelihood monotonically~\cite{Bishop2006}, the convergence of parameter estimation is guaranteed.
The estimated $t$-distribution parameters are finally transformed to variance distribution parameters using $\alpha = \nu/2$ and $\beta = \nu s/2$.
Using this procedure, the variance distribution can be estimated from measured EMG signals.

\section{Experiments}
\subsection{Simulation}
To evaluate the accuracy of the estimated variance distribution, we performed
a simulation experiment using artificially generated EMG signals.
First, a discrete series $\{\sigma^2_t; t=1, \cdots, T\}$ was produced from random numbers following an inverse gamma distribution ${\rm IG}(\alpha_0, \beta_0)$, and a random number $\{x_t\}$ following a Gaussian distribution with a mean of zero and a variance of $\sigma_t$ was then generated for each value of $t$.
The $\{x_t\}$ values were regarded as a time series of an EMG signal measured at a sampling frequency of $F_s$ Hz.
The accuracy of the distribution estimation was verified by comparing the true values $\alpha_0$ and $\beta_0$ with the estimated values $\alpha$ and $\beta$.
As an index of the estimation accuracy, the absolute percentage error was defined as $|{\rm true\ value} - {\rm estimated\ value}|/({\rm true\ value}) \times 100$.

In the estimation of $\alpha$ and $\beta$, the last $L$ values of the signals $\{x_t\}$ were used.
The window length $L$ took values of $100$, $50$, $10$, $5$, $2$, and $1$ s.
The average absolute percentage errors were calculated by changing the true values $400$ times ($\alpha_0 = 0.5, 1.0, 1.5, \cdots, 10.0$, $\beta_0 = 0.05, 0.10, 0.15, \cdots, 1.00$).
To examine the estimation accuracy with respect to variations in the true values $\alpha_0$ and $\beta_0$, the average absolute percentage errors were also calculated for each value of $\alpha_0$ and $\beta_0$ with a fixed value of $L = 5$ s.
The $T$ and $F_{\rm s}$ values in this experiment were set as 100 s and 2,000 Hz.
The initial parameter values in estimation were set using uniform random numbers in the range of (0,~50), and the convergence threshold was set to $\varepsilon = 10^{-7}$.

\subsection{EMG Analysis}
To analyze the variance distribution observed with changes in muscle force, estimation experiments for the EMG variance distribution were conducted using FDI and BB muscle.
Nine healthy young adults (males, right-handed, age range: 22--25 years; mean age: 23.4$\pm$0.9) were voluntarily recruited to conduct an independent task for each muscle.
All participants were in good general health with no self-reports of physical, neurological, or sensory disorders and no history of intense exercise in the previous 24 hours.
They were told the aim of the study and provided written informed consent before participating in the trial.
The experiments were approved by the Hiroshima University Ethics Committee (Registration number: E-840).

\begin{figure}[!t]
\centering
\includegraphics[width=0.95\hsize]{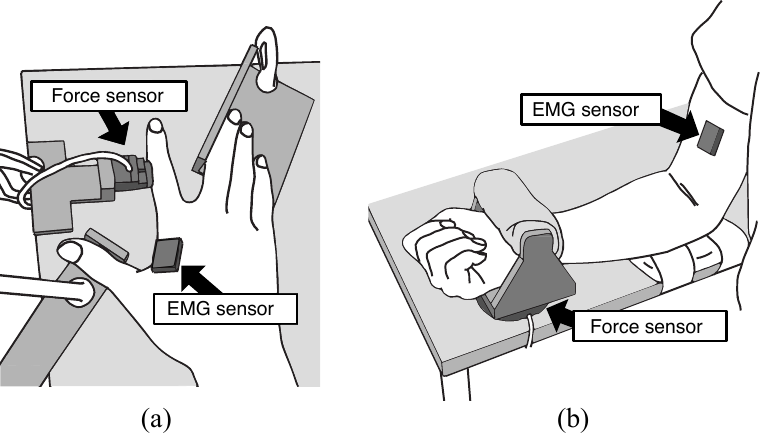}
\caption{Scenes of the EMG recording.
(a) FDI task. Participants were seated with the right upper arm pointing downward, the right shoulder was abducted at $45^{\circ}$ and the elbow was flexed at $90^{\circ}$.
(b) BB task. The right elbow was flexed at $90^{\circ}$ and rested on a padded surface elevated from the table.}
\label{fig:exp_scene}
\end{figure}

During the FDI task, participants were seated with their right upper arm pointing downward, the right shoulder abducted at $45^{\circ}$, and the elbow flexed at $90^{\circ}$.
The right forearm was pronated and immobilized on the desk.
The fingers were also fixed, except for the index finger (Fig.~\ref{fig:exp_scene}(a)).
The abduction force produced by the contraction of the FDI muscle was measured using a force sensor (Leptrino, CFS018CA101U; 16-bit A/D; sampling frequency: 1,200 Hz) installed on the left side of the index finger.
During the BB task, participants were seated with their right upper arm pointing downward, the right lower arm bent forward to the horizontal, and the palm turned upward.
The right elbow was flexed at $90^{\circ}$ and rested on a padded surface elevated from the table (Fig.~\ref{fig:exp_scene}(b)).
The elbow flexion force produced by contraction of the BB muscle was measured by pulling a force sensor (Leptrino, PFS080YA501U6; 16-bit A/D; sampling frequency: 1,200 Hz) fixed on the table with the right wrist.
EMG signals were recorded from the electrode attached on the skin surface of each muscle using a wireless measurement system (Delsys, Trigno; 16-bit A/D; sampling frequency: 2,000 Hz; cutoff frequency: 20--450 Hz).
The experimental protocol was the same for both muscle tasks.

First, each participant performed at least two maximum voluntary contraction (MVC) trials.
The trials were repeated until the peak forces of two MVCs were within 5\% of each other.
Participants rested for at least 2 min between MVC trials.
The greatest peak force of all conducted trials was determined as the MVC force.
Next, participants were instructed to exert constant isometric force at $5\%$, $10\%$, $15\%$, $30\%$, $50\%$, $70\%$, and $80\%$ of MVC force in a random order.
The participants were presented with the exerted force in the form of a bar graph in real time.
The participants were instructed to gradually exert force and increase it to match the target force within 3 s.
When the target was reached, the participants were instructed to maintain that force.
After $5$ s, visual feedback of the exerted force was removed, and the participants were asked to maintain a constant effort for an additional 6 s.
The latter 5 s of the recorded data without visual feedback was used for EMG analysis.
Three trials were conducted for each target force.
A minimum of 1 min of rest was taken between each measurement to avoid muscle fatigue.

The variance distribution parameters [$\alpha, \beta$] were estimated from the measured EMG signals for each target force.
The settings of the initial values and the convergence threshold in parameter estimation were as per the simulation experiment.
A goodness-of-fit test was then conducted to validate the proposed scale mixture model for the real EMG data.
To evaluate the goodness-of-fit, the AD statistic~\cite{Anderson1952} was used.
This measures how well data fit a particular distribution and is calculated as
\begin{equation}
	A^2 = -n-\sum^{n}_{i=1} \frac{2i-1}{n}\left[\ln(F(x_i))+\ln(1-F(x_{n+1-i}))\right]
\end{equation}
where $n$ is the sample size, $\{x_1 < \cdots < x_n\}$ are the ordered sample data, and $F(\cdot)$ is the cumulative distribution function of the specified distribution.
Smaller values of $A^2$ indicate a better fit to the specified distribution.
$A^2$ was calculated by fitting the measured data to the scale mixture model based on the estimated variance distribution.
For comparison, the index was also calculated by fitting the data to a Gaussian distribution model~\cite{Hogan1980c}, a Laplacian distribution model~\cite{Clancy1999}, and the previous model proposed by Hayashi \textit{et al.}~\cite{Hayashi2017}.
The parameters of the Gaussian and Laplacian distribution models were estimated based on maximum-likelihood estimation.
In the previous model, variance distribution parameters were estimated based on an approximation method using rectified/smoothed EMG signals.
This method requires a prior shape parameter $\hat{\alpha}^{\rm pre}$, which is calculated from pre-measured EMG signals $X^{\rm pre}$ and is fixed in advance for each participant.
Accordingly, 10,000 samples were randomly sampled from the measured EMG signals of all trials for the target force of 80\% MVC, and were handled as $X^{\rm pre}$.

\section{Results}
\subsection{Simulation}
\begin{figure}[!t]
\centering
\includegraphics[width=0.85\hsize]{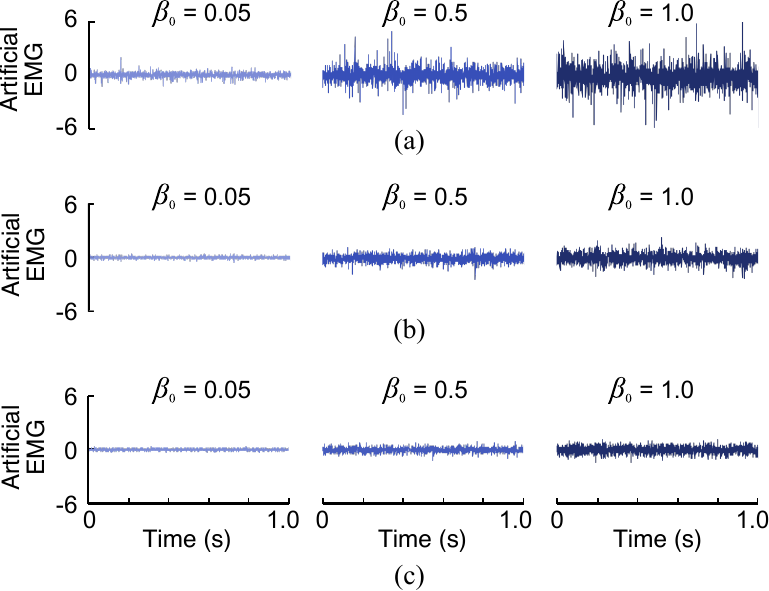}
\caption{Examples of artificial EMG signals with $\alpha_0$ set to (a) $\alpha_0 = 2.5$, (b) $\alpha_0 = 5.0$, and (c) $\alpha_0 = 10.0$; in each case, $\beta_0$ was set to $\beta_0 = 0.05$, $\beta_0 = 0.5$, and $\beta_0 = 1.0$.
}
\label{fig:ex_sim}
\end{figure}

Fig.~\ref{fig:ex_sim} shows time-series waveforms of artificial EMG signals with the data generation parameter $\alpha_0$ set to $2.5$, $5.0$, and $10.0$, and $\beta_0$ changed to $0.01$, $0.10$, and $0.20$ for each value of $\alpha_0$.
%
%
The vertical and horizontal axes indicate the signal values and time, respectively.
For each condition of true values, variance distribution parameters were estimated using a number of initial values selected on the basis of uniform random numbers. In the results, the estimations converged to the same values.

\begin{figure}[!t]
\centering
\includegraphics[width=0.85\hsize]{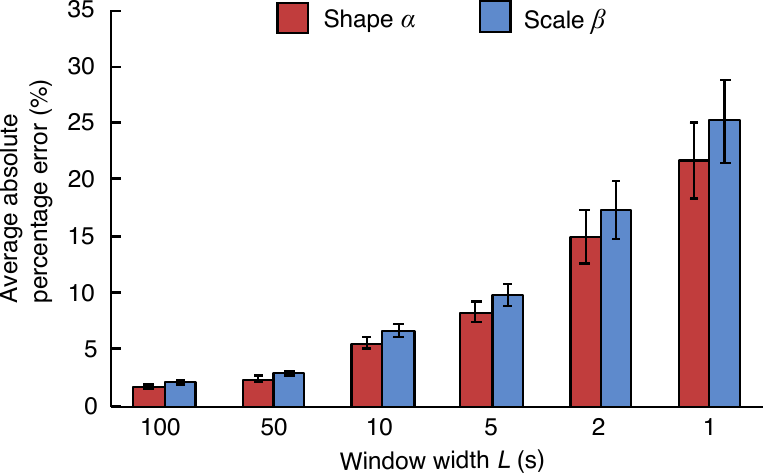}
\caption{Average absolute percentage errors for each window width $L$ in the estimation of $\alpha$ and $\beta$. Error bars represent 95\% confidence intervals for all trials.
}
\label{fig:win_size_sim}
\end{figure}

\begin{figure}[!t]
\centering
\includegraphics[width=0.90\hsize]{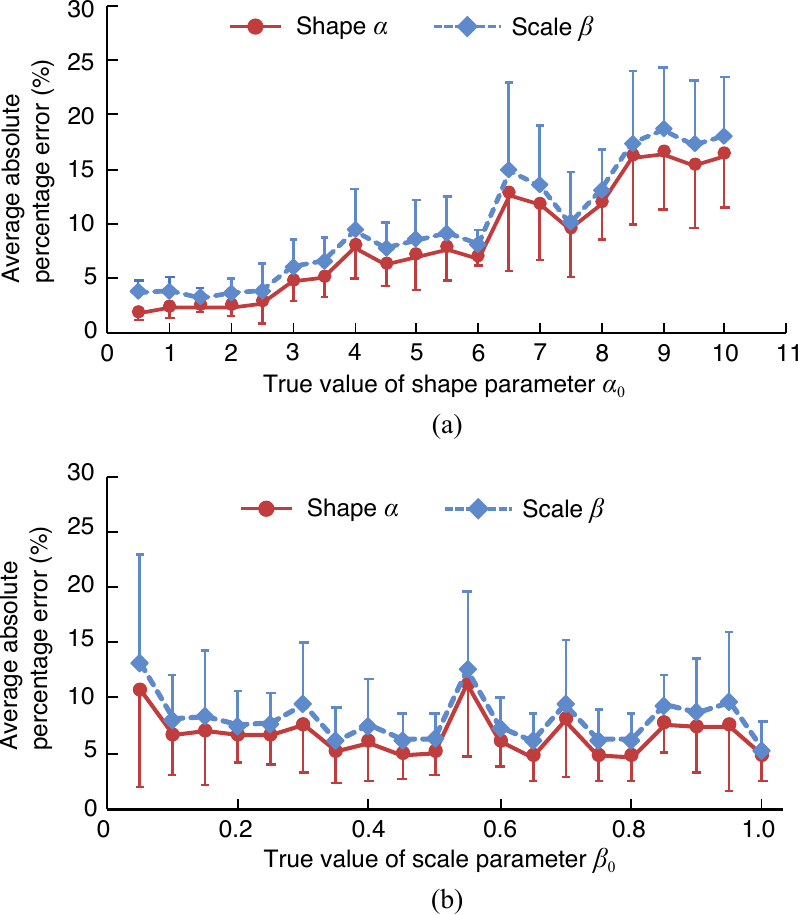}
\caption{Average absolute percentage errors in the estimation of $\alpha$ and $\beta$ for each true value of (a) the shape parameter $\alpha_0$ and (b) the scale parameter $\beta_0$.
Error bars represent 95\% confidence intervals for all trials.
}
\label{fig:true_sim}
\end{figure}

Fig.~\ref{fig:win_size_sim} shows the average absolute percentage errors in the estimation of $\alpha$ and $\beta$ for each value of $L$.
%
%
Fig.~\ref{fig:true_sim} shows the relationship between variations in the true value of the shape and scale parameters [$\alpha_0, \beta_0$] and the estimation accuracy of $\alpha$ and $\beta$ for a fixed value of $L=5$ s.
%
%
The estimation accuracy for $\alpha_0$ and $\beta_0$ was subjected to two-way repeated measures ANOVA (significance level: 0.1\%).
The two factors were the true value and parameter ($\alpha$ and $\beta$).
For $\alpha_0$, both the effect of the true value ($p < 0.0001$) and the effect of the parameter  ($p < 0.0001$) were significant.
For $\beta_0$, only the effect of the parameter was significant ($p < 0.0001$).

\subsection{EMG Analysis}
\begin{figure}[!t]
\centering
\includegraphics[width=0.90\hsize]{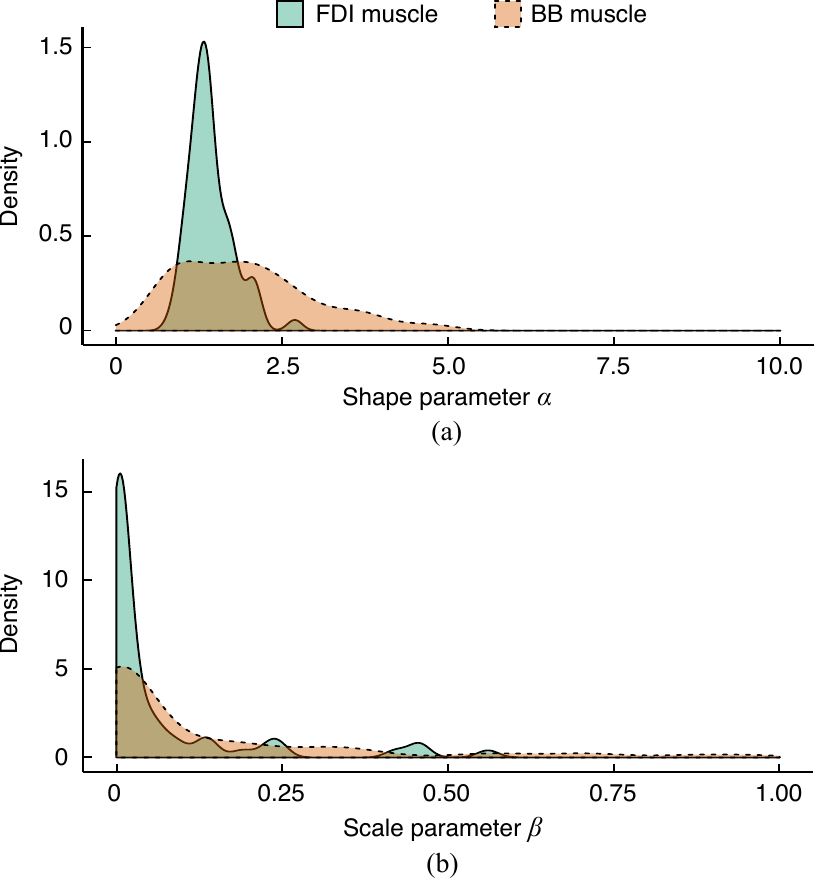}
\caption{Experimental distributions of variance distribution parameters estimated from measured EMG signals for FDI and BB muscle.
(a) Shape parameter $\alpha$. (b) Scale parameter $\beta$.
Distributions are calculated for all participants and all target forces using kernel density estimation.
}
\label{fig:param_dist}
\end{figure}

\begin{figure*}[!t]
\centering
\includegraphics[width=0.95\hsize]{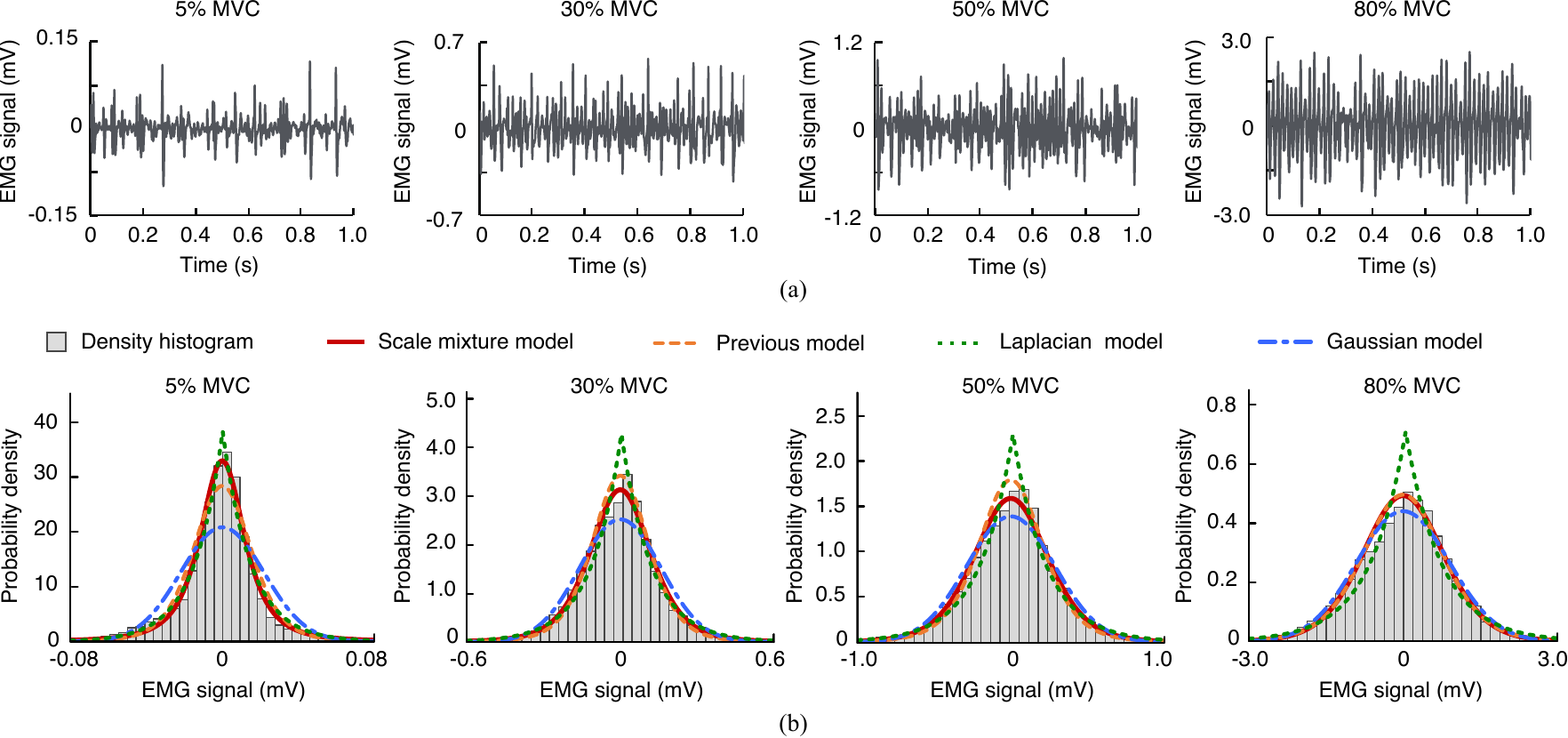}
\caption{Examples of measured EMG signals and corresponding density histograms recorded from BB muscle.
(a) Measured EMG signals recorded from BB muscle for 5, 30, 50, and 80\%\ MVC.
(b) Density histogram of measured EMG signals recorded from BB muscle and fitted distributions based on the proposed scale mixture model, our previous model, the Laplacian model, and the Gaussian model for 5, 30, 50, and 80\%\ MVC.}
\label{fig:exp_wave_dist}
\end{figure*}

\begin{figure}[!t]
\centering
\includegraphics[width=0.95\hsize]{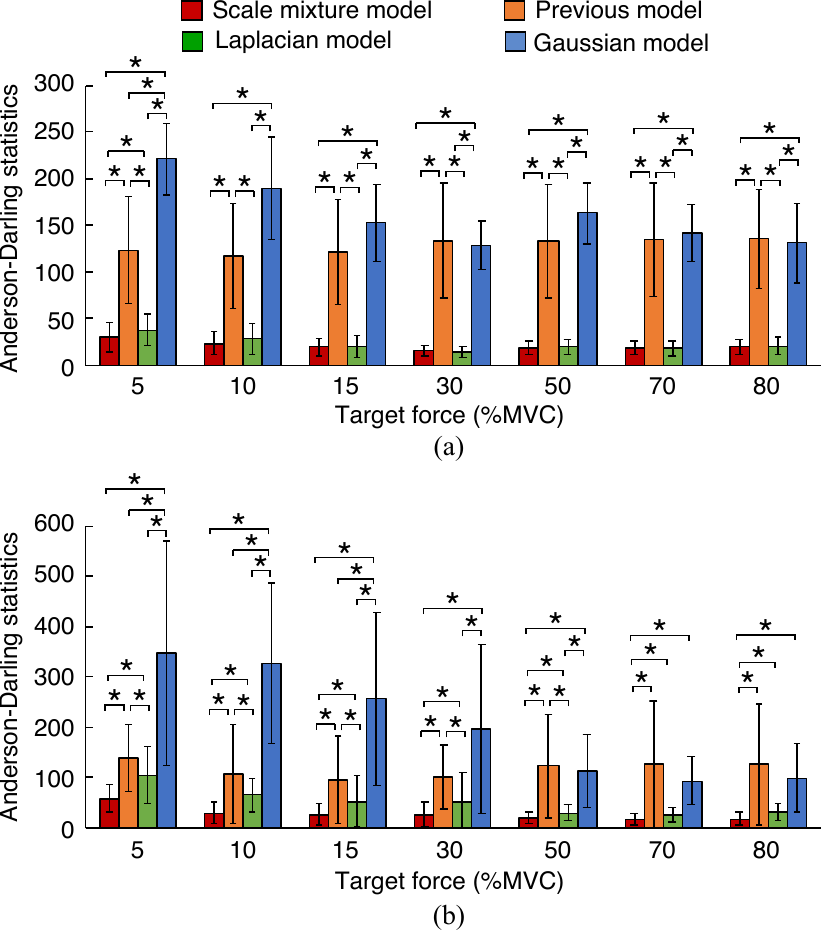}
\caption{Results of goodness-of-fit test for each muscle.
(a) FDI muscle. (b) BB muscle.
Error bars represent 95\% confidence intervals for all participants.
The statistical test results based on the Wilcoxon signed rank test with the Holm-Bonferroni adjustment are also shown (*: $p<0.05$).}
\label{fig:result_gof}
\end{figure}

Fig.~\ref{fig:param_dist} shows the experimental distributions of $\alpha$ and $\beta$ as estimated from measured EMG signals for FDI and BB muscle.
These were calculated for all participants and all target forces using kernel density estimation~\cite{Parzen1962} with a Gaussian kernel.
In this figure, $\alpha$ and $\beta$ are below 4.73 and 0.96, respectively.
%
%
Fig.~\ref{fig:exp_wave_dist}(a) shows examples of measured EMG signals for target forces of 5, 30, 50, and 80\%\ MVC.
%
%
Fig.~\ref{fig:exp_wave_dist}(b) shows density histograms of these measured EMG signals.
Fitted distributions with the scale mixture model, the previous model, the Laplacian model, and the Gaussian model are also shown as solid lines, dashed lines, dotted lines, and dash-dotted lines, respectively.
Fig.~\ref{fig:result_gof} shows the AD statistics for FDI and BB muscle.
%
%
The statistical test results based on the Wilcoxon signed rank test with the Holm-Bonferroni adjustment are also presented.
For FDI muscle, there were significant differences between the scale mixture model and both the previous model and the Gaussian model for all target forces.
A significant difference between the scale mixture model and the Laplacian model was observed only for a target force of 5\% MVC.
For BB muscle, the AD statistics of the scale mixture model were significantly lower than those of the other methods.

\begin{figure*}[!t]
\centering
\includegraphics[width=0.85\hsize]{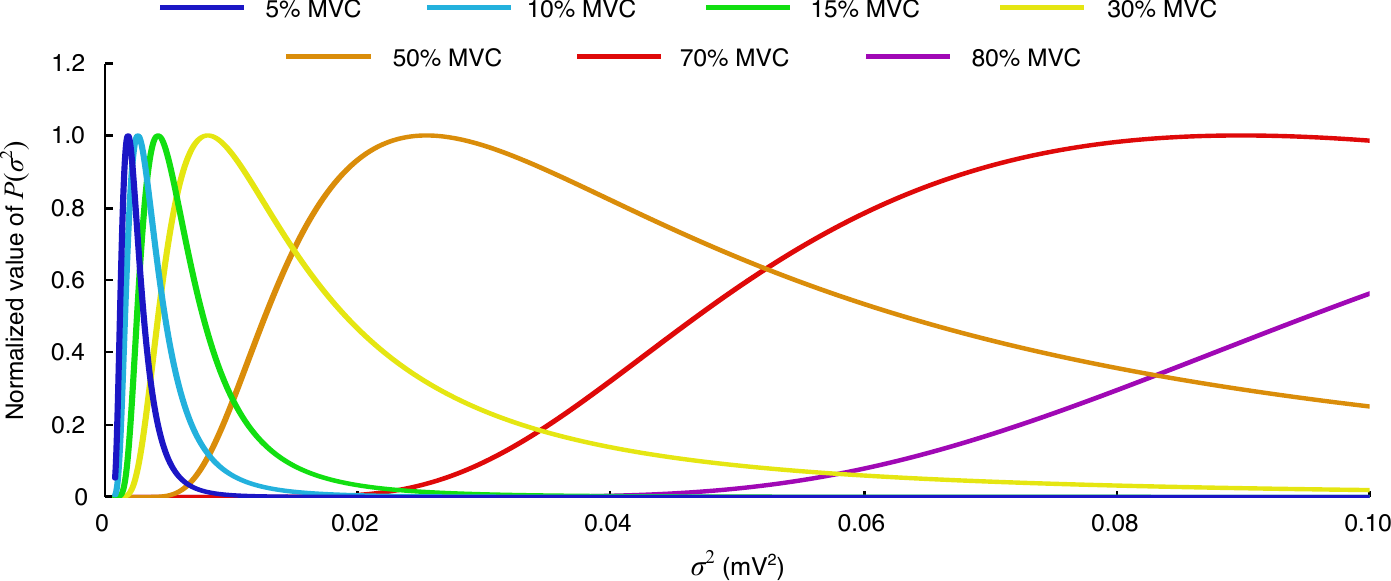}
\caption{Example of variance distribution $P(\sigma^2)$ for measured EMG signals from FDI muscle.
The distributions of all target forces are normalized by their respective peak values of $P(\sigma^2)$ and shown overlapping.}
\label{fig:vd_change}
\end{figure*}

\begin{figure}[!t]
\centering
\includegraphics[width=0.95\hsize]{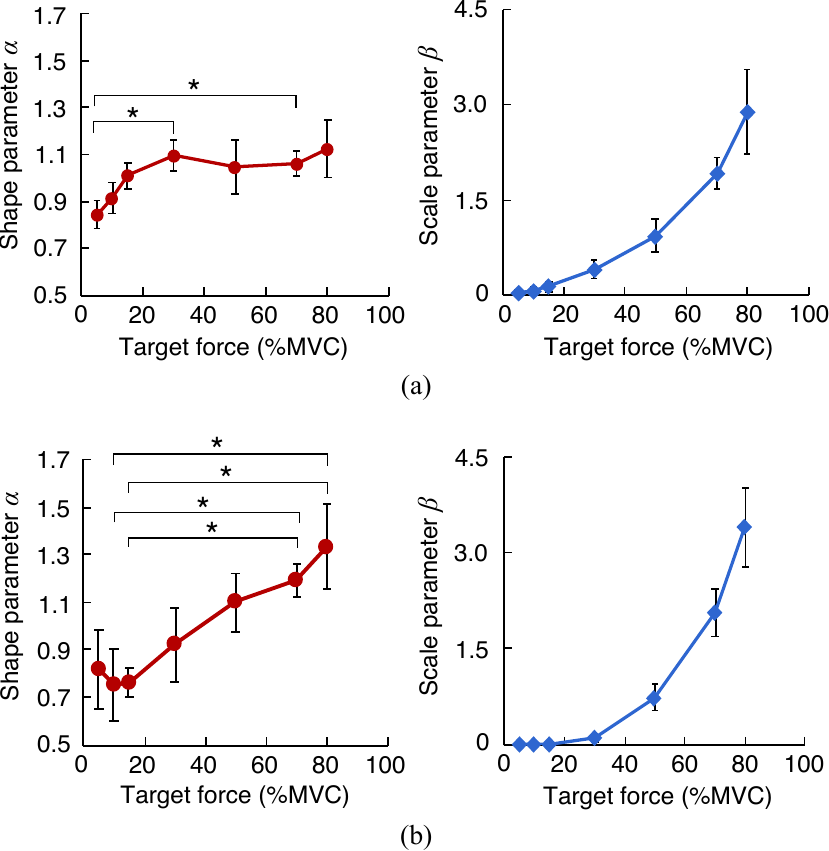}
\caption{Mean of estimated variance distribution parameters $\alpha$ and $\beta$ for each target force.
(a) FDI muscle. (b) BB muscle.
Each estimated parameter was normalized based on division by the mean value for each participant.
Error bars represent 95\% confidence intervals for all participants.
The post hoc test results for $\alpha$ based on the paired two-tailed $t$-test with the Holm-Bonferroni adjustment are also shown (*: $p<0.05$).
}
\label{fig:result_alpha_beta}
\end{figure}

Fig.~\ref{fig:vd_change} shows the example of the estimated variance distribution $P(\sigma^2)$ in the FDI task, with the distributions of all target forces shown overlapping.
%
%
Note that each distribution was normalized based on division by each peak value to adjust the vertical axis scale.
Fig.~\ref{fig:result_alpha_beta} summarizes the mean of $\alpha$ and $\beta$ estimated from FDI muscle and BB muscle in all participants for each target force.
%
%
In this figure, each estimated parameter was normalized to remove individual differences (e.g., skin impedance and skinfold thickness) based on division by the mean value for each participant.
Two-way (muscle and target force) ANOVA (significance level: $5\%$) was conducted for each estimation result of $\alpha$ and $\beta$.
There were significant main effects of the target force for both $\alpha$ ($p < 0.0001$) and $\beta$ ($p < 0.0001$).
In the results for $\alpha$, a significant interaction (muscle $\times$ target force) was observed ($p<0.01$).
Post hoc tests were then conducted for $\alpha$ based on the paired two-tailed $t$-test with the Holm-Bonferroni adjustment; the results are shown in Fig.~\ref{fig:result_alpha_beta}.

\section{Discussion}
In the simulation experiment, the estimation results for $\alpha$ and $\beta$ converged to the same values starting from several different initial values.
Thus, convergence to a global maximum appears to have occurred.
This indicates that the proposed parameter estimation method is robust regardless of the setting of initial values.

In Fig.~\ref{fig:win_size_sim}, the average absolute percentage errors in the estimation of $\alpha$ and $\beta$ at $L=100$ are approximately 2\%, indicating that the variance distribution could be estimated accurately.
The estimation accuracy, however, decreased as the window length $L$ decreased, and the error rate exceeded $24$\% at $L=1$ s, indicating that the estimation accuracy of the variance distribution parameters depends on the estimation window length $L$.
Thus, using more samples will improve the accuracy of the estimation.
In this paper, the variance distribution parameters are estimated by a maximum-likelihood method based on the EM algorithm.
Hence, based on the law of large numbers, a large number of samples are required for accurate estimation.
These results show that at least 5 s of EMG signals are needed to estimate the variance distribution with an error rate less than 10\%.

Fig.~\ref{fig:true_sim} shows that the estimation accuracy of the variance distribution parameters decreases as the true value of the shape parameter $\alpha_0$ increases.
In contrast, the true value of the scale parameter $\beta_0$ does not significantly affect the estimation accuracy.
Here, the shape parameter $\alpha$ corresponds to the degree of freedom $\nu$ in Student's $t$-distribution (recall the change of variable in the derivation of the parameters estimation method).
Therefore, $\nu$ reflects the Gaussianity of the distribution, and the distribution becomes closer to a Gaussian form as $\nu$ increases.
Thus, if the true value of $\alpha$ is large, it may be difficult to discriminate between the true distribution of the data and the Gaussian distribution, resulting in the estimation accuracy of the parameters decreasing.

In the EMG analysis experiment, the variance distribution parameters estimated from the measured EMG signals were within the ranges of the true values used in the simulation experiment (Fig.~\ref{fig:param_dist}).
This result suggests that variance distribution can also be estimated with real EMG signals at an accuracy of approximately 5--10\% as verified by the simulation experiment.

In Fig.~\ref{fig:exp_wave_dist}(b), the density histograms of the measured EMG signals are more peaked near zero than the Gaussian distributions, and their tails are wider.
These differences are notable when the target force is small, in which case the measured EMG signals exhibit characteristics similar to those of the Laplacian distribution, whereas they tend to approach the Gaussian distribution as the target force increases.
These results are in agreement with previous findings reported by Clancy and Hogan~\cite{Clancy1999}.
The scale mixture model successfully reflects the distribution characteristics of the measured EMG signals at all target forces and obtains a better fit.
This can be confirmed from the results of the goodness-of-fit test shown in Fig.~\ref{fig:result_gof}.
The AD statistics for the scale mixture model are significantly lower than those of the previous model proposed by Hayashi \textit{et al.}~\cite{Hayashi2017} and the Gaussian model for all target forces in both muscles.
The scale mixture model also gives a significantly better fit than the Laplacian model at 5\% MVC in FDI muscle and at all target forces in BB muscle.
This is because the scale mixture model considers the variance distribution parameters $\alpha$ and $\beta$ to reflect the variability of EMG variance by introducing the variance distribution and can represent the varying tail weight of the distributions, i.e., the change in non-Gaussianity.
In contrast, the Gaussian model only has a variance parameter, meaning its tail weight is treated as a constant.
The Laplacian model has a heavier tail than the Gaussian model and gives better fitting results than the Gaussian distribution.
However, the Laplacian model cannot represent the change in non-Gaussianity according to muscle activity because it also has a constant tail weight.
The goodness of fit of the previous model fell between those of the Gaussian model and the Laplacian model.
Although the previous model also has a heavier tail than the Gaussian model, its tail weight is constant because the shape parameter $\alpha$ controlling non-Gaussianity is fixed in advance.
Consequently, the previous model showed worse fitting results than the proposed model.
This suggests that accurate evaluation of changes in EMG signal distribution shape requires estimation of both $\alpha$ and $\beta$, which is achievable with the proposed model.
The previous model based on approximate estimation using rectified/smoothed EMG signals is therefore effective only when the actual shape parameter is known.

Based on the above, it can be concluded that the scale mixture model enables better fitting to real EMG signals than the previous model~\cite{Hayashi2017} and other conventional stochastic EMG models.
In previous studies reporting that the probability distribution of EMG signals falls between a Gaussian and a Laplacian distribution~\cite{Clancy1999, Nazarpour2013}, it was difficult to identify the switching point of these distributions because the non-Gaussianity of such signals continuously varies with muscle force.
By contrast, the proposed scale mixture model enables description of changes in the non-Gaussianity of EMG signals associated with muscle activity within the unified scheme.
Further, accurate modeling of the stochastic properties of EMG signals is important in extracting the features of such signals and quantitatively evaluating neuromuscular activity.
Recent studies have suggested that the non-Gaussianity of EMG signals serves as a potentially important index for EMG pattern classification~\cite{Nazarpour2007}, muscular-activity detection~\cite{Meigal2009}, and evaluation of neuromuscular disease~\cite{Orosco2015}.
Accordingly, higher performance may be realized in application fields such as prosthesis control if novel EMG processing techniques can be developed using the scale mixture model.

In Fig.~\ref{fig:vd_change}, the variance distributions $P(\sigma^2)$ shift to the right and spread horizontally as the muscle force increases, indicating that the absolute dispersion of EMG variance increases in response to increased muscle force.
Such a change in variance distribution shape is caused by the combination of changes in $\alpha$ and $\beta$~\cite{Furui2017a}.
In Fig.~\ref{fig:result_alpha_beta}, both $\alpha$ and $\beta$ increase as the target force increases in the FDI and BB muscles.
The scale parameter $\beta$ shows exponential increases in both muscles, and there are no significant differences between FDI muscle and BB muscle.
As $\beta$ determines the spread of the variance distribution, this may be influenced by increases in EMG variance (i.e., amplitude) according to muscle force.
Although $\alpha$ also increases with the target force, it exhibits a different tendency in FDI muscle than in BB muscle.
This is also suggested by the significant interaction (muscle $\times$ target force) in the results for $\alpha$.
This intermuscular difference in the variance distribution parameter $\alpha$ is a novel finding, as experiments in previous studies were performed only for BB muscle~\cite{Hayashi2017,Furui2017a}.

Note that $\alpha$ determines the Gaussianity of the EMG distribution, as described above.
Thus, the changes in $\alpha$ shown in Fig.~\ref{fig:result_alpha_beta} indicate changes in the Gaussianity of the EMG signals, and the distribution of the EMG signals approaches the Gaussian distribution as the muscle force increases.
This is a result of the central limit theorem (CLT), whereby the EMG signals become more Gaussian as the muscle force increases because of the larger number of (independent) motor units firing~\cite{Milner-Brown1975, Nazarpour2013, Mcgill2004}.
In contrast, if the muscle force is small, only a small number of motor units are recruited, and the EMG signals are the summation of a sparse pattern of motor units firing~\cite{Stegeman2016}.
In this case, the individual motor unit firings can be considered discrete events separated by time periods.
Hence, the EMG signals have outlier-like variance, resulting in the non-Gaussian and non-stationary features of EMG signals.
Such changes in EMG signals can also be seen from the features of the measured waveform shown in Fig.~\ref{fig:exp_wave_dist}(a).

Based on the CLT hypothesis, the different behavior of $\alpha$ between FDI muscle and BB muscle may be attributed to the difference in the recruitment strategies of the motor units in each muscle.
In the case of FDI muscle, the upper limit of motor unit recruitment is approximately 50--75\% MVC~\cite{Moritz2005a}.
Hence, the change in the Gaussianity of the EMG signals is expected to occur up to approximately 60\% MVC.
In fact, the increases in $\alpha$ in FDI muscle are only confirmed at relatively low target forces, and $\alpha$ remains almost constant for target forces above 50\% MVC (Fig.~\ref{fig:result_alpha_beta}(a)).
In contrast, the upper limit of motor unit recruitment in BB muscle is known to be at least 85\% MVC~\cite{Kukulka1981a}.
Hence, it is expected that the Gaussianity of the EMG signals will change up to approximately this MVC.
Fig.~\ref{fig:result_alpha_beta}(b) shows that the value of $\alpha$ in BB muscle increases over a wide range of target forces and continues to increase linearly up to 80\% MVC.
Therefore, the change in the Gaussianity of the measured EMG signals, as characterized by $\alpha$, is related to the recruitment patterns in motor units, as has been experimentally reported in previous studies.
From the above observations, the variance distribution parameter $\alpha$ in the scale mixture model may reflect the changes in the number of recruited motor units according to the muscle force.

\section{Conclusion}
In this paper, we modeled surface EMG signals using a scale mixture model with a variance distribution.
In the model, the EMG variance is handled as a random variable that follows an inverse gamma distribution, and the probability distribution of EMG signals is assumed to be an infinite mixture of Gaussians having the same mean but different variances.
The scale mixture model involves the assumption that the non-Gaussianity of EMG signals is caused by stochastic fluctuations in EMG variance.
This helps to clarify changes in the non-Gaussianity of EMG signals associated with muscle activity.

Simulations using artificially generated EMG signals revealed that the estimation accuracy of the variance distribution depends on the number of samples used for the estimation.
The EMG analysis experiment using FDI and BB muscle revealed that the scale mixture model provides a better fit to the measured EMG signals than conventional EMG models.
Moreover, this experiment shows that the variance distribution parameters may reflect underlying motor unit activity based on the relationship between the Gaussianity of the EMG signal distribution and the muscle force.

In the measurement experiments conducted in this study, EMG signals were measured from healthy participants under constant-force and non-fatigue conditions.
However, the firings and recruitment patterns of motor units are changed by physiological factors such as neural disorders and fatigue~\cite{Contessa2009a,GLENDINNING1994,Zwarts2000}; hence, these factors may influence the EMG variance distribution and its parameters.
Therefore, in future research, the authors plan to investigate the characteristics of EMG variance distributions in relation to other muscle conditions (e.g., muscular fatigue).
In addition, we will analyze the relationship between the recruitment pattern of motor units and the variance distribution parameters in detail through the simultaneous measurement of surface and intramuscular EMG signals.


%



\ifCLASSOPTIONcaptionsoff
  \newpage
\fi



%


\begin{thebibliography}{10}
	\providecommand{\url}[1]{#1}
	\csname url@samestyle\endcsname
	\providecommand{\newblock}{\relax}
	\providecommand{\bibinfo}[2]{#2}
	\providecommand{\BIBentrySTDinterwordspacing}{\spaceskip=0pt\relax}
	\providecommand{\BIBentryALTinterwordstretchfactor}{4}
	\providecommand{\BIBentryALTinterwordspacing}{\spaceskip=\fontdimen2\font plus
	\BIBentryALTinterwordstretchfactor\fontdimen3\font minus
	  \fontdimen4\font\relax}
	\providecommand{\BIBforeignlanguage}[2]{{%
	\expandafter\ifx\csname l@#1\endcsname\relax
	\typeout{** WARNING: IEEEtran.bst: No hyphenation pattern has been}%
	\typeout{** loaded for the language `#1'. Using the pattern for}%
	\typeout{** the default language instead.}%
	\else
	\language=\csname l@#1\endcsname
	\fi
	#2}}
	\providecommand{\BIBdecl}{\relax}
	\BIBdecl
	
	\bibitem{Moynes1986}
	D.~R. Moynes \emph{et~al.}, ``{Electromyography and motion analysis of the
	  upper extremity in sports},'' \emph{Phys. Ther.}, vol.~66, no.~12, pp.
	  1905--1911, Dec. 1986.
	
	\bibitem{Millington1992}
	P.~J. Millington \emph{et~al.}, ``{Biomechanical analysis of the sit-to-stand
	  motion in elderly persons.}'' \emph{Arch. Phys. Med. Rehabil.}, vol.~73,
	  no.~7, pp. 609--617, Jul. 1992.
	
	\bibitem{tao2012}
	W.~Tao \emph{et~al.}, ``{Gait analysis using wearable sensors},''
	  \emph{Sensors}, vol.~12, no.~2, pp. 2255--2283, Feb. 2012.
	
	\bibitem{Baweja2009}
	H.~S. Baweja \emph{et~al.}, ``{Removal of visual feedback alters muscle
	  activity and reduces force variability during constant isometric
	  contractions},'' \emph{Exp. Brain Res.}, vol. 197, no.~1, pp. 35--47, Jul.
	  2009.
	
	\bibitem{Ganguly2011}
	K.~Ganguly \emph{et~al.}, ``{Reversible large-scale modification of cortical
	  networks during neuroprosthetic control},'' \emph{Nat. Neurosci.}, vol.~14,
	  no.~5, pp. 662--669, May 2011.
	
	\bibitem{Busche2015}
	M.~A. Busche \emph{et~al.}, ``{Rescue of long-range circuit dysfunction in
	  Alzheimer's disease models},'' \emph{Nat. Neurosci.}, vol.~18, no.~11, pp.
	  1623--1630, Nov. 2015.
	
	\bibitem{Sturman2005}
	M.~M. Sturman \emph{et~al.}, ``{Effects of aging on the regularity of
	  physiological tremor},'' \emph{J. Neurophysiol.}, vol.~93, no.~6, pp.
	  3064--3074, Jun. 2005.
	
	\bibitem{Cifrek2009}
	M.~Cifrek \emph{et~al.}, ``{Surface EMG based muscle fatigue evaluation in
	  biomechanics},'' \emph{Clin. Biomech.}, vol.~24, no.~4, pp. 327--340, May
	  2009.
	
	\bibitem{Dideriksen2011}
	J.~L. Dideriksen \emph{et~al.}, ``{EMG-based characterization of pathological
	  tremor using the iterated hilbert transform},'' \emph{IEEE Trans. Biomed.
	  Eng.}, vol.~58, no.~10, pp. 2911--2921, Oct. 2011.
	
	\bibitem{Fukuda2003}
	O.~Fukuda \emph{et~al.}, ``{A human-assisting manipulator teleoperated by EMG
	  signals and arm motions},'' \emph{IEEE Trans. Robot. Autom.}, vol.~19, no.~2,
	  pp. 210--222, Apr. 2003.
	
	\bibitem{Scheme2011}
	E.~Scheme and K.~Englehart, ``{Electromyogram pattern recognition for control
	  of powered upper-limb prostheses: State of the art and challenges for
	  clinical use},'' \emph{J. Rehabil. Res. Dev.}, vol.~48, no.~6, pp. 643--660,
	  2011.
	
	\bibitem{Farina2017}
	D.~Farina \emph{et~al.}, ``{Man/machine interface based on the discharge
	  timings of spinal motor neurons after targeted muscle reinnervation},''
	  \emph{Nat. Biomed. Eng.}, vol.~1, no.~2, p. 0025, Feb. 2017.
	
	\bibitem{Parker1977}
	P.~Parker \emph{et~al.}, ``{Signal processing for the multistate myoelectric
	  channel},'' \emph{Proc. IEEE}, vol.~65, no.~5, pp. 662--674, 1977.
	
	\bibitem{Hogan1980c}
	N.~Hogan and R.~W. Mann, ``{Myoelectric signal processing: Optimal estimation
	  applied to electromyography---Part I: Derivation of the optimal
	  myoprocessor},'' \emph{IEEE Trans. Biomed. Eng.}, vol. BME-27, no.~7, pp.
	  382--395, Jul. 1980.
	
	\bibitem{Abbink1998}
	J.~H. Abbink \emph{et~al.}, ``{Detection of onset and termination of muscle
	  activity in surface electromyograms.}'' \emph{J. Oral Rehabil.}, vol.~25,
	  no.~5, pp. 365--369, May 1998.
	
	\bibitem{Milner-Brown1975}
	H.~S. Milner-Brown and R.~B. Stein, ``{The relation between the surface
	  electromyogram and muscular force.}'' \emph{J. Physiol.}, vol. 246, no.~3,
	  pp. 549--569, Apr. 1975.
	
	\bibitem{Hunter1987}
	I.~W. Hunter \emph{et~al.}, ``{Estimation of the conduction velocity of muscle
	  action potentials using phase and impulse response function techniques},''
	  \emph{Med. Biol. Eng. Comput.}, vol.~25, no.~2, pp. 121--126, Mar. 1987.
	
	\bibitem{Clancy1999}
	E.~A. Clancy and N.~Hogan, ``{Probability density of the surface electromyogram
	  and its relation to amplitude detectors},'' \emph{IEEE Trans. Biomed. Eng.},
	  vol.~46, no.~6, pp. 730--739, Jun. 1999.
	
	\bibitem{Bilodeau1997}
	M.~Bilodeau \emph{et~al.}, ``{Normality and stationarity of EMG signals of
	  elbow flexor muscles during ramp and step isometric contractions},'' \emph{J.
	  Electromyogr. Kinesiol.}, vol.~7, no.~2, pp. 87--96, 1997.
	
	\bibitem{Naik2011}
	G.~R. Naik \emph{et~al.}, ``{Kurtosis and negentropy investigation of myo
	  electric signals during different MVCs},'' in \emph{Proc. ISSNIP Biosignals
	  Biorobotics Conf. 2011}, Jan. 2011, pp. 1--4.
	
	\bibitem{Nazarpour2013}
	K.~Nazarpour \emph{et~al.}, ``{A note on the probability distribution function
	  of the surface electromyogram signal},'' \emph{Brain Res. Bull.}, vol.~90,
	  pp. 88--91, Jan. 2013.
	
	\bibitem{Zhao2012}
	Y.~Zhao and D.~Li, ``{A simulation study on the relation between muscle motor
	  unit numbers and the non-Gaussianity/non-linearity levels of surface
	  electromyography},'' \emph{Sci. China Life Sci.}, vol.~55, no.~11, pp.
	  958--967, Nov. 2012.
	
	\bibitem{Messaoudi2017}
	N.~Messaoudi \emph{et~al.}, ``{Assessment of the non-Gaussianity and
	  non-linearity levels of simulated sEMG signals on stationary segments},''
	  \emph{J. Electromyogr. Kinesiol.}, vol.~32, pp. 70--82, Feb. 2017.
	
	\bibitem{Hayashi2017}
	H.~Hayashi \emph{et~al.}, ``{A variance distribution model of surface EMG
	  signals based on inverse gamma distribution},'' \emph{IEEE Trans. Biomed.
	  Eng.}, vol.~64, no.~11, pp. 2672--2681, Nov. 2017.
	
	\bibitem{Furui2017a}
	A.~Furui \emph{et~al.}, ``{Variance distribution analysis of surface EMG
	  signals based on marginal maximum likelihood estimation},'' in \emph{Proc.
	  Eng. Med. Biol. Soc. (EMBC), 2017 39th Annu. Int. Conf. IEEE}, Jeju Island,
	  Jul. 2017, pp. 2514--2517.
	
	\bibitem{Bishop2006}
	C.~M. Bishop, \emph{{Pattern recognition and machine learning}}.\hskip 1em plus
	  0.5em minus 0.4em\relax Springer-Verlag New York, Inc., 2006.
	
	\bibitem{Anderson1952}
	T.~W. Anderson and D.~A. Darling, ``{Asymptotic theory of certain ``goodness of
	  fit" criteria based on stochastic processes},'' \emph{Ann. Math. Stat.},
	  vol.~23, no.~2, pp. 193--212, Jun. 1952.
	
	\bibitem{Parzen1962}
	E.~Parzen, ``{On estimation of a probability density function and mode},''
	  \emph{Ann. Math. Stat.}, vol.~33, no.~3, pp. 1065--1076, Sep. 1962.
	
	\bibitem{Nazarpour2007}
	K.~Nazarpour \emph{et~al.}, ``{Application of higher order statistics to
	  surface electromyogram signal classification},'' \emph{IEEE Trans. Biomed.
	  Eng.}, vol.~54, no.~10, pp. 1762--1769, Oct. 2007.
	
	\bibitem{Meigal2009}
	A.~Meigal \emph{et~al.}, ``{Novel parameters of surface EMG in patients with
	  Parkinson's disease and healthy young and old controls},'' \emph{J.
	  Electromyogr. Kinesiol.}, vol.~19, no.~3, pp. e206--e213, Jun. 2009.
	
	\bibitem{Orosco2015}
	E.~Orosco \emph{et~al.}, ``{On the use of high-order cumulant and bispectrum
	  for muscular-activity detection},'' \emph{Biomed. Signal Process. Control},
	  vol.~18, pp. 325--333, Apr. 2015.
	
	\bibitem{Mcgill2004}
	K.~C. Mcgill, ``{Surface electromyogram signal modeling},'' \emph{Med. Biol.
	  Eng. Comput.}, vol.~42, no.~4, pp. 446--454, 2004.
	
	\bibitem{Stegeman2016}
	D.~F. Stegeman \emph{et~al.}, ``{EMG modeling and simulation},'' in
	  \emph{Electromyogr. Physiol. Eng. Noninvasive Appl.}\hskip 1em plus 0.5em
	  minus 0.4em\relax Hoboken, NJ, USA: John Wiley {\&} Sons, Inc., Jan. 2005,
	  pp. 205--231.
	
	\bibitem{Moritz2005a}
	C.~T. Moritz \emph{et~al.}, ``{Discharge rate variability influences the
	  variation in force fluctuations across the working range of a hand muscle},''
	  \emph{J. Neurophysiol.}, vol.~93, no.~5, pp. 2449--2459, May 2005.
	
	\bibitem{Kukulka1981a}
	C.~G. Kukulka and H.~P. Clamann, ``{Comparison of the recruitment and discharge
	  properties of motor units in human brachial biceps and adductor pollicis
	  during isometric contractions},'' \emph{Brain Res.}, vol. 219, no.~1, pp.
	  45--55, 1981.
	
	\bibitem{Contessa2009a}
	P.~Contessa \emph{et~al.}, ``{Motor unit control and force fluctuation during
	  fatigue},'' \emph{J. Appl. Physiol.}, vol. 107, no.~1, pp. 235--243, Jul.
	  2009.
	
	\bibitem{GLENDINNING1994}
	D.~S. Glendinning and R.~M. Enoka, ``{Motor unit behavior in Parkinson's
	  disease.}'' \emph{Phys. Ther.}, vol.~74, no.~1, pp. 61--70, Jan. 1994.
	
	\bibitem{Zwarts2000}
	M.~J. Zwarts \emph{et~al.}, ``{Recent progress in the diagnostic use of surface
	  EMG for neurological diseases},'' \emph{J. Electromyogr. Kinesiol.}, vol.~10,
	  no.~5, pp. 287--291, Oct. 2000.
	
	\end{thebibliography}




\end{document}